\documentclass{egs}                  

\usepackage{times}                   
\usepackage{graphicx}
\usepackage{amssymb}

\printfigures              

\begin{document}
\sloppy
\title{Fractional Fourier approximations for potential gravity waves
on deep water}

 \author{Vasyl P. Lukomsky$^1$}
 \author{Ivan S. Gandzha$^2$}
 \affil{Institute of Physics, Kyiv, Ukraine}
 \affil{Electronic address: $^1$\texttt{lukom@iop.kiev.ua}, $^2$\texttt{gandzha@iop.kiev.ua}}

\date{Manuscript version from \today}
\journal{\NPG}       

 \firstauthor{Gandzha}
 \proofs{V. P. Lukomsky\\Department of
Theoretical Physics, Institute of Physics\\ Prospect Nauky 46,
Kyiv 03028, Ukraine}
 \offsets{V. P. Lukomsky\\Department of
Theoretical Physics, Institute of Physics\\ Prospect Nauky 46,
Kyiv 03028, Ukraine}

\msnumber{}
 \received{??}
 \revised{??} 
 \accepted{??}
 \runninghead{Lukomsky and Gandzha: Fractional Fourier approximations}
 \firstpage{1} \pubyear{} \pubvol{} \pubnum{}

\maketitle

\begin{abstract}
In the framework of the canonical model of hydrodynamics, where
fluid is assumed to be ideal and incompressible, waves are
potential, two-dimensional, and symmetric, the authors have
recently reported the existence of a new type of gravity waves on
deep water besides well studied Stokes waves \citep{prl2002}. The
distinctive feature of these waves is that horizontal water
velocities in the wave crests exceed the speed of the crests
themselves. Such waves were found to describe irregular flows with
stagnation point inside the flow domain and discontinuous
streamlines near the wave crests. Irregular flows produce a simple
model for describing the initial stage of the formation of
spilling breakers when a localized jet is formed at the crest
following by generating whitecaps.

In the present work, a new highly efficient method for computing
steady potential gravity waves on deep water is proposed to
examine the above results in more detail. The method is based on
the truncated fractional approximations for the velocity potential
in terms of the basis functions
$1/\bigr(1-\exp(y_0-y-ix)\bigl)^n$, $y_0$ being a free parameter.
The non-linear transformation of the horizontal scale $x = \chi -
\gamma \sin\chi,\; 0<\gamma<1,$ is additionally applied to
concentrate a numerical emphasis on the crest region of a wave for
accelerating the convergence of the series. Fractional
approximations were employed for calculating both steep Stokes
waves and irregular flows. For lesser computational time, the
advantage in accuracy over ordinary Fourier expansions in terms
the basis functions $\exp\bigl(n (y+ix)\bigr)$ was found to be
from one to ten decimal orders depending on the wave steepness and
flow parameters.
\end{abstract}

\section{Introduction}\label{sec:intro}


From old times the wave motion of the ocean bewitched and
extremely attracted the attention of mankind. Up to now the
problem of understanding specific features of water waves and
their modelling represent a real challenge both from scientific
and engineering points of view. Occurrence of extremely large and
steep ocean breaking waves imposes a hazard to fishing boats,
ships, and off-shore oil facilities. To understand physical
mechanisms that give rise to extreme breaking waves and to model
them correctly it is necessary to gain detailed knowledge of the
form and dynamics of steep water waves.

The canonical problem about the propagation of surface waves on
deep water (see Section \ref{sec:model}) was the first essentially
non-linear problem in hydrodynamics. Its analysis during almost
two hundred years gave the origin to many fields of non-linear
dynamics such as solitary waves, modulation instabilities, strange
attractors, etc. \citet{Stokes1847} was the first who considered
surface waves of finite amplitude (Stokes waves). \citet{Stokes}
conjectured that such waves must have a maximal amplitude (the
limiting wave) and showed the flow in this wave to be singular at
the crest forming a $120^{\circ}$ corner (the Stokes corner flow).
Much later, \citet{Grant} suggested that this singularity, for a
wave that has not attained the limiting form, is located above the
wave crest and forms a stagnation point with streamlines meeting
at right angles. \citet{Long:al-h1978} proved this numerically
after extending Stokes flows analytically outside the domain
filled by fluid. The following question resulted: why the flow in
the limiting Stokes wave has the $120^{\circ}$ singularity instead
of the $90^{\circ}$ one, as in any wave with lesser amplitude?
Because of this \citet{Grant} conjectured that a continuous
approach to the limiting amplitude is possible only if the Stokes
corner flow has several coalescing singularities. However, it has
not yet clear where these multiple singularities arise from.

A new era in developing the theory of steep gravity waves started
from the work of \citet{Long:1975}, where he found that many
characteristics of gravity waves, such as speed, energy, and
momentum, are not monotonic functions of the wave amplitude, as
was assumed from Stokes, but attain total maxima and then drop
before the limiting wave is reached. \citet{Long:al-h1977}
constructed asymptotic expansions for waves close to the
$120^{\circ}$-cusped wave (almost highest waves) and showed that
these dependences oscillate infinitely as the limiting wave is
approached. Nevertheless, strict numerical verification of such
oscillations seems to be a real challenge up to the present time,
with only the first relative maximum and minimum having been
thoroughly investigated \citep[see, e.g.,][]{Long-Tanaka}.

\citet{Tanaka:1983} showed that gravity waves steeper than the
wave with maximal total energy become unstable with respect to
two-dimensional disturbances having the same period as an
undisturbed wave (superharmonic instability). \citet{Jillians}
investigated the form of such instabilities and showed that they
lead to wave overturning and breaking. The conjecture was made
that wave breaking is a purely local phenomenon around the wave
crest which, in the case of spilling breakers and more gently
plunging breakers, occurs independently of the flow in the rest of
a wave. Proceeding with this idea \citet{Long:al-h1994-1} and
\citet{Long:al-h1994-2} suggested that superharmonic instability
results in the crests of almost highest Stokes waves to be
unstable (crest instability). \citet{Long-Tanaka} strongly
supported the conclusion that superharmonic instabilities of
Stokes waves are indeed crest instabilities. Finally,
\citet{Long:al-h1997} showed that crest instabilities lead (i) to
wave overturning and breaking or (ii) to a smooth transition of a
wave to a lower progressive wave having nearly the same total
energy, followed by a return to a wave of almost the initial wave
height. The latter fact generated a new question: what is the
nature of such a transient phenomenon? A possible explanation
would be found if superharmonic instability resulted in a
bifurcation to a new solution, as usually takes place in
non-linear dynamics. However, up to this time it was assumed that
the Stokes solution is unique and free of bifurcations in keeping
with the uniqueness argument of \citet{Garabedian}. The only
bifurcation known to occur is the trivial one of a pure phase
shift at the point of energy maximum \citep{Tanaka:1985}.

The above results are all related to Stokes waves, for which the
speed of fluid particles at the wave crests is smaller than the
wave phase speed, equality being achieved for the limiting wave
only. Thus, the traditional criterion for wave breaking is that
horizontal water velocities in the crest must exceed the speed of
the crest \citep{Peregrine:review}. \citet{cpc2002, prl2002} have
recently provided evidence (although numerical and not completely
rigorous) for the existence of a new family of two-dimensional
irrotational symmetric periodic gravity waves that satisfy the
criterion of breaking. A stagnation point in the flow field of
these waves is inside the flow domain, in contrast to the Stokes
waves of the same wavelength. This makes streamlines exhibit
discontinuity in the vicinity of the wave crests, with
near-surface particles being jetted out from the flow. Because of
this such waves and flows were called irregular (in contrast to
regular Stokes flows). Existence of irregular waves made the
authors propose the following conjectures \citep[see][]{prl2002}:
(i) a $120^{\circ}$ corner at the crest of the limiting Stokes
wave seems to be formed by merging the stagnation points of a
regular Stokes flow and an irregular flow; (ii) irregular flows
are apparently the result of developing superharmonic or crest
instabilities of regular Stokes flows.

To calculate irregular flows \citet{cpc2002, prl2002} used
truncated Fourier expansions for the velocity potential and the
elevation of a free surface in the plane of spatial variables (a
physical plane). Debiane and Kharif
\citep[see][]{abstr:EGS02_infinite} confirmed the existence of
irregular waves using inverse plane Longuet-Higgins method
\citep{Long:1986}, where the spatial coordinates are represented
as Fourier series in velocity potential and stream function, the
corresponding coefficients being evaluated by solving quadratic
relations between them. Finally, \citet{Clamond:rKdV} also
obtained irregular flows by applying his new renormalized cnoidal
wave (RCW) approximation. Note, however, that at present these are
only numerical and approximate proofs.

Ordinary Fourier expansions used by \citet{cpc2002} become not
efficient enough for approximating irregular waves and even Stokes
waves close to the limiting one due to slow descending of the
Fourier coefficients. Thus, a more efficient method is necessary
for these tasks. Up to now the most precise and efficient way for
calculating the properties of two-dimensional surface waves is
Tanaka's method of the inverse plane
\citep[see][]{Tanaka:1983,Tanaka:1986}. The key idea of his method
is to map the inverse plane into a unit circle by means of the
Nekrasov transformation. Then boundary conditions are transformed
to an integral equation, which is solved iteratively. The accuracy
of obtained solutions is drastically improved by concentrating a
numerical emphasis on the crest region using further
transformation of variables. As a result, Tanaka's method is the
only one being capable of evaluating the second maximum of the
phase speed and even further higher order extremums. In spite of
all the advantages of Tanaka's method and his program, where it is
implemented, we are interested in improving the methods of the
physical plane since they can be applied for calculating 3D waves
as well and can be generalized to the case of non-ideal and
compressible fluid, in contrast to all the inverse plane methods.

Thus, the purpose of this paper is to present a new method in the
physical plane for calculating two-dimensional potential steady
progressive surface waves on the fluid of infinite depth (see
Section \ref{sec:method}). The method is based on the fractional
Fourier approximation for the velocity potential recently
introduced by the authors \citep{abstr:EGS02_infinite,abstr:CCP02}
and the non-linear transformation of the horizontal scale for
concentrating a numerical emphasis near the wave crest. The first
term of such a fractional Fourier approximation was independently
derived by \citet{Clamond:rKdV} and was called a renormalized
cnoidal wave approximation.

In Section \ref{sec:results}, fractional Fourier approximations
are applied for calculating regular and irregular flows. In
Section \ref{sec:Stokes}, great numerical advantage of fractional
approximations over ordinary Fourier approximations is
demonstrated when calculating almost highest Stokes waves.
Although the accuracy of the results is still less than the ones
obtained using Tanaka's method, proposed fractional Fourier
approximations have a potential to become almost as effective as
the method of Tanaka. In Section \ref{sec:irregular}, fractional
approximations allowed investigating the character of singularity
of irregular flows in more detail. The profiles of irregular waves
are shown to reveal the Gibbs phenomenon usually taking place when
a discontinuous function or a continuous function with
discontinuous derivatives are approximated by continuous truncated
Fourier series. Moreover, even regular Stokes waves very close to
the Stokes corner flow are also demonstrated to exhibit the
similar Gibbs phenomenon in accordance with the observation of
\citet{Nekrasov_Gibbs}. Concluding remarks are given in Section
\ref{sec:concl}.

\section{The canonical model}\label{sec:model}
\begin{figure}[t]
\includegraphics[width=8.5cm]{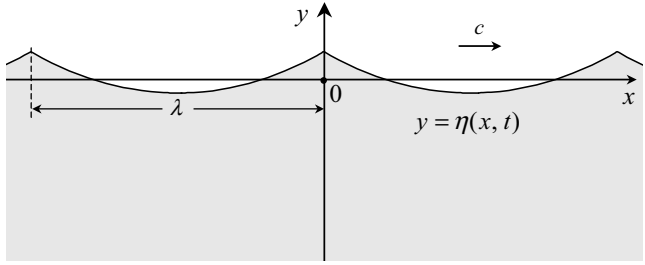}
\caption{\label{fig:water} The laboratory frame of reference.}
\end{figure}

Consider the dynamics of potential two-dimensional periodic waves
on the irrotational, inviscid, incompressible fluid with unknown
free surface under the influence of gravity. Waves are assumed to
propagate without changing their form from left to right along the
$x$-axis with constant speed $c$ relative to the motionless fluid
at infinite depth (see Fig.~\ref{fig:water}). Gravity waves and
related fluid flows are governed by the following set of equations
\begin{eqnarray}
\label{eq:Lapl}
 \Phi_{\theta\theta}+\Phi_{yy}=0, && -\infty<y<\eta(\theta);
\\ \label{eq:Dyn}
 \left(c-\Phi_{\theta}\right)^2+\Phi_y^2+2\eta=c^2,
 && y=\eta(\theta);
\\ \label{eq:Kin}
 \left(c-\Phi_{\theta}\right)\eta_{\theta}+\Phi_y=0, && y=\eta(\theta);
\\ \label{eq:Inf}
 \Phi_\theta=0,~\Phi_y=0, && y=-\infty;
\end{eqnarray}
where $\theta=x-ct$ is the wave phase, $\Phi(\theta,~y)$ is the
velocity potential (the velocity is equal to
$\overrightarrow{\nabla}\Phi$), $\eta(\theta)$ is the elevation of
the unknown free surface, and $y$ is the upward vertical axis such
that $y=0$ is the still water level. Herein (\ref{eq:Lapl}) is the
Laplace equation in the flow domain, (\ref{eq:Dyn}) is the
dynamical boundary condition (the Bernoulli equation),
(\ref{eq:Kin}) is the kinematic boundary condition (no fluid
crosses the surface), (\ref{eq:Inf}) is the condition that fluid
is motionless at infinite depth. The dimensionless variables are
chosen such that length and time are normalized by the wavenumber
$k$ and the frequency $\sqrt{gk}$ of a linear wave, respectively,
$g$ being the acceleration due to gravity. In this case, the
dimensionless wavelength $\lambda = 2\pi$.

Once the velocity potential and the wave phase speed are known,
particle trajectories (streamlines) are found from the following
differential equations:
\begin{equation}\label{eq:traject}
 \frac{\mathrm{d} \theta}{\mathrm{d} t}=\Phi_\theta\left(\theta,~y\right)-c,
 \quad \frac{\mathrm{d} y}{\mathrm{d} t}=\Phi_y\left(\theta,~y\right);
\end{equation}
in the frame of reference moving together with the wave (the wave
related frame of reference). In this frame of reference, each
streamline is characterized by a constant value of a stream
function $\psi(\theta,~y)$. The velocity potential and the stream
function $\Psi(\theta,~y) = \psi(\theta,~y) + cy$ in the
laboratory frame of reference are connected by means of the
Cauchy-Riemann conditions:
\begin{equation}\label{eq:Cauchy}
\Phi_\theta = \Psi_y; \; \Phi_y = -\Psi_\theta.
\end{equation}
This makes possible introducing the complex potential
$W=\Phi+i\Psi$ so that
\begin{equation} \label{eq:R}
\Phi=-i c (R-R^*),~~\Psi = c (R+R^*),~~R_\theta=i R_y;
\end{equation}
where $R = iW^*/2c$, $^*$ is the complex conjugate. In terms of
the complex function $R(\theta,~y)$, the dynamical and kinematic
boundary conditions (\ref{eq:Dyn}), (\ref{eq:Kin}) are as follows:
\begin{eqnarray}
\label{eq:DynR}
 i \left(R_\theta - R_\theta^*\right)+2R_\theta R_\theta^*+\eta/c^2=0,
 && y=\eta(\theta);
\\ \label{eq:KinR}
 R(\theta,~\eta)+R^*(\theta,~\eta)-\eta=B, && B = \mathrm{const};
\end{eqnarray}
where $B$ is the Bernoulli constant. It is found by averaging
(\ref{eq:KinR}): $B = K - \overline{\eta}$,~~$K = I/c$ is the
Stokes flow, $I$ is the wave impulse averaged over the period:
$$
I = \frac{1}{2\pi} \int\limits_{0}^{2\pi} \mathrm{d}\theta
\int\limits_{-\infty}^{\eta(\theta)}
\Phi_\theta(\theta,~y)\mathrm{d}y =
\frac{1}{2\pi}\int\limits_{0}^{2\pi}
\Psi\left(\theta,~\eta(\theta)\right)\mathrm{d}\theta .
$$

In addition to the Laplace equation and the boundary conditions,
an initial condition should be assigned. Since the canonical model
is energy conservative, the wave total energy can be used instead
to characterise wave properties. For this purpose, however, the
crest-to-trough height $H$ or the wave steepness $A=H/\lambda$ are
the more convenient parameters since they monotonously increase
starting from linear waves up to the limiting configuration. Thus,
using the Eqs. (\ref{eq:Lapl}), (\ref{eq:DynR}), (\ref{eq:KinR}),
(\ref{eq:Inf}) of the canonical model the following quantities are
to be found as the functions of the wave steepness $A$: the
complex function $R(\theta,~y)$, the elevation $\eta(\theta)$ of
the free surface, and the wave phase speed $c$.

\section{The method for obtaining solutions} \label{sec:method}

\subsection{\textit{Fractional Fourier approximations}} \label{sec:y0}
When working in the plane of spatial variables the solutions to
the Laplace equation (\ref{eq:Lapl}) in the flow domain are
usually looked for as the following truncated Fourier series

\begin{equation} \label{eq:Fourier}
R(\theta,~y) = \sum_{n=1}^N \xi_n \exp \bigl(n (y + i
\theta)\bigr).
\end{equation}
This approach was applied by the authors \citep{cpc2002, prl2002}
for calculating steep gravity waves. Fourier expansions
(\ref{eq:Fourier}), however, become ineffective for steep waves
with sharpening crests close to the limiting wave due to slow
descending of the Fourier coefficients. Because of this we
proposed \citep{abstr:EGS02_infinite, abstr:CCP02} a more
effective set of functions to expand the velocity potential on the
basis of the following Euler formula \citep[see][]{Hamming}
\begin{eqnarray}\label{eq:Euler}
\sum\limits_{n=1}^{\infty}\sigma_n~z^n &=&
\sum\limits_{n=1}^{\infty}\frac{\zeta_n}{\left(1-z^{-1}\right)^n},
\nonumber\\
&&\zeta_n =
\sum\limits_{n_1=1}^{n}(-1)^{n_1}C_{n-1}^{n_1-1}\sigma_{n_1},
\end{eqnarray}
$C_n^{n_1}$ being the binomial coefficients. By choosing
$z(\theta,~y)=\exp\left(y-y_0+i\theta\right)$; $\sigma_n=\xi_n
\exp\left(n y_0\right)$ the following one-parametric expansion for
the velocity potential is obtained after truncating the series:
\begin{eqnarray}
R(\theta,~y;~y_0) &=& \sum_{n=1}^N \frac{\zeta_n }{\bigl(1 - \exp
\left(y_0 - y - i\theta \right)\bigr)^n} \nonumber \\
&=& \sum_{n=1}^N \frac{\alpha_n }{\bigl(\exp(-y_0) - \exp \left(-
y - i\theta \right)\bigr)^n} \nonumber\\
&\equiv& \sum_{n=1}^{N}\alpha_n T^n(\theta,~y;~y_0), \label{eq:R_y0}\\
T(\theta,~y;~y_0)&=&\bigl(\exp(-y_0) - \exp \left(- y - i\theta
\right)\bigr)^{-1}, \label{eq:T}
\end{eqnarray}
where the normalized coefficients $\alpha_n = \zeta_n \exp(-n
y_0)$ were introduced to overcome infinite exponents at
$y_0\rightarrow \infty$. Approximation (\ref{eq:R_y0}) shows a
formal correspondence with Pad\'e-type fractional approximates.
Because of this we called expansion (\ref{eq:R_y0}) a ``fractional
Fourier expansion". It is singular in a countable number of
isolated points $y = y_0$, $\theta=2\pi k,~k\in \mathbb{Z}$, their
location being determined by a free parameter $y_0$. Singular
points are to be located outside the flow domain for calculating
potential waves. At $y_0 = \infty$, fractional Fourier expansion
(\ref{eq:R_y0}) reduces to ordinary Fourier expansion
(\ref{eq:Fourier}) with $\xi_n=(-1)^n \alpha_n$. Due to
(\ref{eq:Euler}) expansions (\ref{eq:Fourier}) and (\ref{eq:R_y0})
are equivalent at $N = \infty$ and the convergence of
(\ref{eq:R_y0}) follows from the convergence of
(\ref{eq:Fourier}). For finite $N$ and $y_0 \sim 1$, however, a
fractional Fourier expansion converges much more rapidly than an
ordinary Fourier expansion. The reason is that a finite number of
terms in (\ref{eq:R_y0}) always corresponds to an infinite number
of terms in (\ref{eq:Fourier}) that is especially important for
sharp-crested waves.

Expansions (\ref{eq:R_y0}), (\ref{eq:Fourier}), and, in general,
any function $R(\theta,~y) = R(y+i\theta)$ all satisfy the Laplace
equation (\ref{eq:Lapl}) exactly. The latter fact was also used by
\citet{Clamond:renorm, Clamond:rKdV} (for finite and infinite
depth, respectively) to introduce a renormalization principle that
allows reconstructing the velocity potential in the whole domain
once the velocity potential at the bottom (or any other level) is
known. By applying such renormalization to the first-order
periodic solution of KdV equation \citet{Clamond:rKdV} obtained
the velocity potential being exactly the same to the first term
($N=1$) of expansion (\ref{eq:R_y0}), which he called a
renormalized cnoidal wave (RCW) approximation. There may be other
possibilities to improve ordinary Fourier expansion
(\ref{eq:Fourier}) besides the proposed fractional expansion
(\ref{eq:R_y0}). However, one should additionally assure the
convergence of series that makes constructing such generalized
expansions much more difficult.

One can see from the expansion of derivatives
\begin{eqnarray} \label{eq:Ry_y0}
R_y(\theta,~y;~y_0) &=& -iR_{\theta} = \sum_{n=1}^{N+1} \beta_n T^n(\theta,~y;~y_0), \nonumber \\
\beta_n &=& n\alpha_n - (n-1)\alpha_{n-1} \exp{(-y_0)},
\end{eqnarray}
which follows directly from (\ref{eq:R_y0}), that the boundary
condition at infinite depth (\ref{eq:Inf}) is also satisfied
exactly.

Hereafter, only the symmetric waves are considered. In this case,
the coefficients $\alpha_n$ and $\xi_n$ are real (in general, they
are complex numbers for nonsymmetric waves). After taking into
account expansions (\ref{eq:R_y0}) and (\ref{eq:Ry_y0}) the
boundary conditions (\ref{eq:DynR}), (\ref{eq:KinR}) at the free
surface attain the following form:
\begin{eqnarray}
\lefteqn{2c^2\Bigl(\sum_{n_1=1}^{N+1}\beta_{n_1} \mathrm{Re}(T^{n_1})-\phantom{\Bigr)}} \nonumber\\
\lefteqn{-\phantom{\Bigl(}\sum_{n_1=1}^{N+1}\sum_{n_2=1}^{N+1}\beta_{n_1}\beta_{n_2}\mathrm{Re}(T^{n_1}T^{*~n_2})\Bigr)
=\eta,~~y=\eta(\theta) \label{eq:Dyn_T};} \\
&&2\sum_{n_1=1}^{N}\alpha_{n_1}\mathrm{Re}(
T^{n_1})-\eta=B,~~y=\eta(\theta). \label{eq:Kin_T}
\end{eqnarray}

Note that Eqs. (\ref{eq:Dyn_T}) and (\ref{eq:Kin_T}) at
$N\rightarrow\infty$ are equivalent to boundary conditions
(\ref{eq:DynR}) and (\ref{eq:KinR}), in the class of
$2\pi$-periodic functions (subharmonic waves with multiple periods
are not taken into account in expansions (\ref{eq:R_y0})).

\subsection{\textit{Nonlinear transformation of the horizontal scale}}\label{sec:gamma}
To solve boundary conditions
(\ref{eq:Dyn_T}) and (\ref{eq:Kin_T}) one should assign an
appropriate approximation to the unknown elevation
$y=\eta(\theta)$ of the free surface. In the plain of spatial
variables, the Fourier series
\begin{equation} \label{eq:Fourier_Eta}
\eta\left(\theta\right)=\sum_{n=-\infty}^{\infty}\eta_n\,\exp(in\theta),~~\eta_{-n}=\eta_n,
\end{equation}
are often used. Taking infinite bounds is not necessary for
profiles with rounded enough crests. However, adequate description
of sharp-crested profiles close to the limiting one requires
taking into account excessively large number of modes due to
extremely slow descending of the Fourier coefficients. This highly
restricts practical application of (\ref{eq:Fourier_Eta}). The
following non-linear transformation of the horizontal scale
originally suggested by \citet{Saffman}
\begin{equation} \label{eq:gamma}
\theta(\chi;~\gamma)=\chi-\gamma\sin\chi,~~0<\gamma<1,
\end{equation}
allows overcoming this difficulty by stretching sharp crests to a
more rounded configuration. As a result, the Fourier series
\begin{equation} \label{eq:Fourier_Eta_gamma}
\eta\left(\chi;~\gamma\right)=\sum_{n=-M}^{M}\eta^{(\gamma)}_n\,\exp(in\chi),~~\eta^{(\gamma)}_{-n}=\eta^{(\gamma)}_n,
\end{equation}
in the $\chi$-space with stretched crests are much more efficient
\citep{abstr:CCP02}. Due to nonlinear transformation
(\ref{eq:gamma}) any finite number $M$ of the coefficients
$\eta^{(\gamma)}_n$ at $\gamma\neq 0$ corresponds to infinite
number of the coefficients $\eta_n\equiv\eta^{(0)}_n$ in ordinary
Fourier series ($\gamma=0$), the associated relations being
presented in Appendix \ref{apex:gamma}. Thus, the role of the
parameter $\gamma$ for the series (\ref{eq:Fourier_Eta}) in
horizontal coordinate $\theta$ is the same to the role of the
parameter $y_0$ for the series (\ref{eq:Fourier}) in vertical
coordinate $y$.

\subsection{\textit{Numerical procedure}} \label{sec:eq}
By means of (\ref{eq:Fourier_Eta_gamma}) the boundary conditions
(\ref{eq:Dyn_T}), (\ref{eq:Kin_T}) at the free surface are reduced
to the following system of non-linear algebraic equations
\begin{eqnarray}
\mathcal{D}_n=c^2 d_n - \eta^{(\gamma)}_n = 0,&& n=\overline{0,~M}; \label{eq:Dyn_n} \\
\mathcal{K}_n=2\sum_{n_1=1}^{N}\alpha_{n_1}
t_n^{(n_1)}-\eta^{(\gamma)}_n = 0 ,&& n=\overline{1,~N};
\label{eq:Kin_n}
\end{eqnarray}
where
$$
d_n=2\sum_{n_1=1}^{N+1}\beta_{n_1} \Bigl(
t_n^{(n_1)}-\sum_{n_2=n_1}^{N+1}\beta_{n_2}(2-\delta_{n_1,~n_2})\,t_n^{(n_1,~n_2)}\Bigr),
$$
$\delta_{n_1,~n_2}$ is the Kronecker delta. The coefficients
$t_n^{(n_1)}$ and $t_n^{(n_1,~n_2)}$ are the Fourier harmonics of
the functions $\mathrm{Re}(T^{n_1})$ and
$\mathrm{Re}(T^{n_1}T^{*~n_2})$, respectively:
\begin{eqnarray}
\lefteqn{\quad~\, t_n^{(n_1)}=\frac{1}{2\pi}\int\limits_0^{2\pi}
\mathrm{Re}\Bigl(T^{n_1}\bigl(\theta(\chi),~\eta(\chi)\bigr)\Bigr)
\exp(-in\chi) \mathrm{d}\chi,} \nonumber\\
\lefteqn{t_n^{(n_1,~n_2)}=\frac{1}{2\pi}\int\limits_0^{2\pi}
\mathrm{Re}\bigl(T^{n_1}T^{*~n_2}\bigr) \exp(-in\chi)
\mathrm{d}\chi.} \label{eq:FFT}
\end{eqnarray}
They were calculated using the fast Fourier transform (FFT). The
Bernoulli constant is found from the kinematic equations
(\ref{eq:Kin_n}) at $n=0$:
\begin{equation} \label{eq:B}
  B = 2\sum_{n_1=1}^{N}\alpha_{n_1} t_0^{(n_1)}-\eta^{(\gamma)}_0.
\end{equation}

The truncation of Eqs. (\ref{eq:Dyn_n}), (\ref{eq:Kin_n}) was
chosen for the following reasons. Since the set of kinematic
equations (\ref{eq:Kin_n}) is linear over the coefficients
$\alpha_n$ ($n=\overline{1,~N}$), they can be found in terms of
the harmonics $\eta^{(\gamma)}_n$ without using dynamical
equations (\ref{eq:Dyn_n}). To proceed in such a way, it is
sufficient to take into account only the first $N$ kinematic
equations. Then the rest $M+1$ variables $c$, $\eta^{(\gamma)}_0$,
$\eta^{(\gamma)}_n$ ($n=\overline{2,~M}$) are found from dynamical
equations (\ref{eq:Dyn_n}). The last unknown parameter
$\eta^{(\gamma)}_1$ is determined by the wave steepness
$A=\bigl(\eta(0)-\eta(\pi)\bigr)/2\pi$ as follows
\begin{equation} \label{eq:A}
\eta^{(\gamma)}_1=\frac{\pi}{2}A-\sum_{n=1}^{[(M-1)/2]}\eta^{(\gamma)}_{2n+1},
\end{equation}
the square brackets designating the integer part. Since the wave
steepness $A$ is an integral characteristic, some wave properties
may be missed when using it as a governing parameter. Thus, we
additionally use the first harmonic $\eta_1$ of the elevation in
the $\theta$-space (a spectral characteristic) as an independent
variable instead of the wave steepness. In this case, the first
harmonic $\eta^{(\gamma)}_1$ in the $\chi$-space is expressed in
terms of $\eta_1$ and the rest of the harmonics
$\eta^{(\gamma)}_n$ ($n=\overline{2,\;M}$) by means of the
expression (\ref{eq:Bessel}) at $n=1$ instead of (\ref{eq:A}).

The set of Eqs. (\ref{eq:Dyn_n}), (\ref{eq:Kin_n}) was solved by
Newton's method, the Jacoby matrix being given in Appendix
\ref{apex:Jacoby}. Starting values for new calculations were taken
from previous runs. For large enough $N$ and $M$, the Jacoby
matrix was found to become badly conditioned. Because of this the
program realization was implemented in arbitrary precision
computer arithmetic. For instance, computations at $N=150$,
$M=2.5N$ demand 160-digit arithmetic that is ten times more
accurate than the machine one. Note that such a run is equivalent
in computer time to a run with $N=250$, $M=4N$ using ordinary
Fourier approximations and because of this takes approximately $4$
times lesser computer memory.

The truncation numbers $N$ and $M$ are chosen for the following
reasons. By fixing the number $N$ in expansion (\ref{eq:R_y0}) an
approximate configuration of the velocity potential is assigned.
To find out a proper truncation of the series
(\ref{eq:Fourier_Eta_gamma}) for the elevation associated with
this configuration, the number $M$ should be increased until the
revision of solutions for greater $M$ becomes less than chosen
accuracy. Then the precision to which boundary conditions
(\ref{eq:KinR}) and (\ref{eq:DynR}) are satisfied defines the
absolute errors connected with a truncation of the potential and
elevation, respectively. The overall relative error
$Er_{\mathrm{max}}$ of such an approximate solution is agreed to
be the maximal absolute error in boundary conditions
(\ref{eq:DynR}) and (\ref{eq:KinR}) all over the wave period
divided by the Bernoulli constant $B$. To obtain a solution close
to the exact one, one should gradually increase the number $N$,
choosing every time a proper value of $M$, until overall desired
precision is achieved. In a majority of calculations, it was
sufficient to use the approximation $M=2.5N$ or lesser ones.

The numerical scheme proposed operates with two parameters $y_0$
and $\gamma$. Decreasing $y_0$ from $y_0=\infty$ to $y_0\sim1$
accelerates the convergence of the fractional Fourier expansion
(\ref{eq:R_y0}) for the velocity potential, lesser $N$ being
necessary to retain the same accuracy. Increasing $\gamma$ from
$\gamma=0$ to $\gamma=1-\varepsilon,~\varepsilon\rightarrow 0$
accelerates the convergence of the expansion
(\ref{eq:Fourier_Eta_gamma}) for the elevation, lesser $M$ being
necessary to retain the same accuracy. These two processes,
however, should be carried out simultaneously. Using the
fractional Fourier expansion (\ref{eq:R_y0}) without the
transformation of the horizontal scale (\ref{eq:gamma}) was found
to deteriorate the convergence of series (\ref{eq:Fourier_Eta})
and, vice versa, using the transformation of the horizontal scale
without the fractional Fourier expansion was found to deteriorate
the convergence of expansion (\ref{eq:Fourier}), with only slight
overall benefit having been achieved. On the contrary, using the
fractional Fourier expansion in combination with the nonlinear
transformation of the horizontal scale proved to be highly
efficient (see Section \ref{sec:results}).

\begin{figure*}[!]
\includegraphics[width=12.5cm]{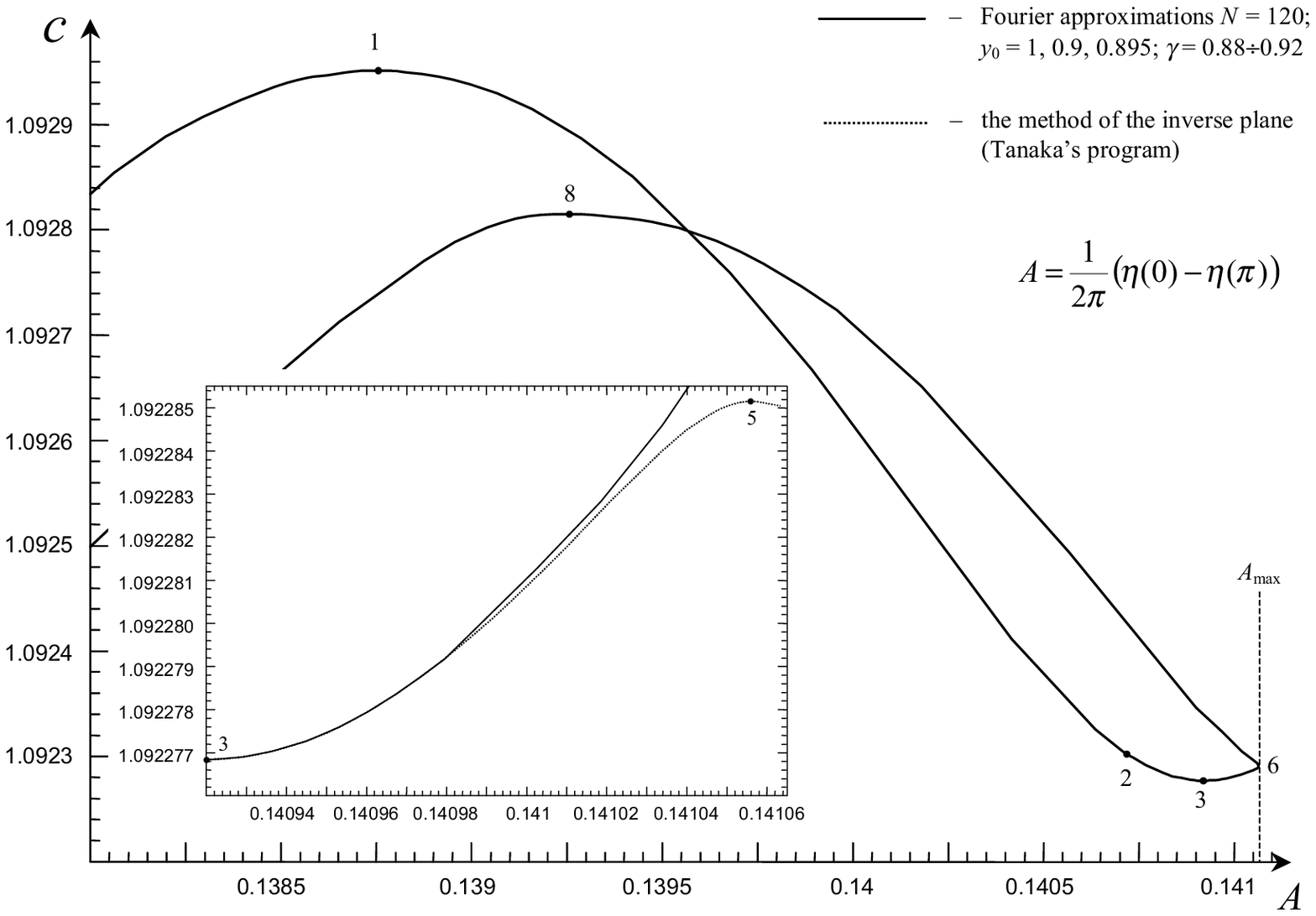}
\caption{\label{fig:c(A)} The dependence of the phase speed $c$ of
steep surface waves on their steepness $A$.} \vspace{5mm}
\includegraphics[width=10cm]{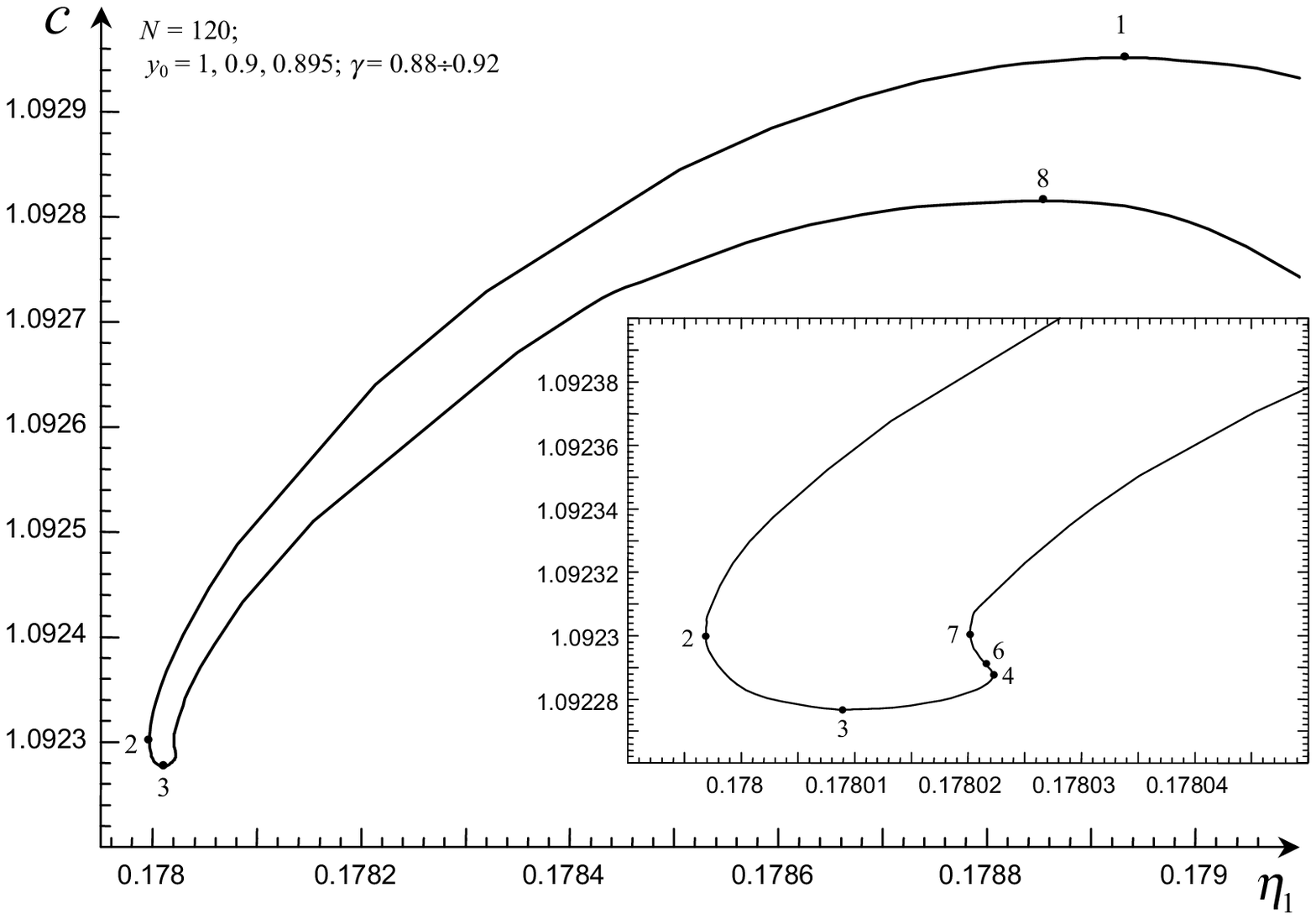}
\caption{\label{fig:c(eta1)} The dependence of the phase speed $c$
of steep surface waves on the first harmonic $\eta_1$ of their
profile.}
\end{figure*}
\begin{table*}[!]
  \caption[]{\label{tab:extremum} The points of extremums in phase speed $c$, steepness $A$ ($\varepsilon=\pi A$),
  and the first harmonic $\eta_1$ \\of the profile for Stokes waves ($N=120,~M=2.5N,~y_0=0.9,~\gamma=0.92$)}
\smallskip
\renewcommand{\arraystretch}{1.2}
\iftwocol{\small}{}
\begin{tabular}{ccclcl}
\hline \hline
 \multicolumn{1}{c}{point} & \multicolumn{1}{c}{extremum} & \multicolumn{1}{c}{$A$} & \multicolumn{1}{c}{$\varepsilon$}&
 \multicolumn{1}{c}{$c$} & \multicolumn{1}{c}{$\eta_1$} \\
 \hline
   & the first max of $\eta_1$ & 0.1351 & 0.424429 & 1.0909437483 & 0.1799822 \\[3pt]
 1 & the first max of $c$ & 0.13875 & 0.435896 & 1.0929513818 & 0.1789318 \\[3pt]
 2 & the first min of $\eta_1$ & 0.14072 & 0.442085 & 1.0923021558 & 0.1779969 \\[3pt]
 3 & the first min of $c$ & 0.14092 & 0.442713 & 1.0922768392 & 0.1780099 \\[3pt]
 4 & the second max of $\eta_1$ & $\approx 0.141055$ & 0.443137 & $\approx1.092288$ & 0.1780222 \\[3pt]
 5$^*$ & the second max of $c$ & 0.141056 & 0.443141 & 1.0922851495 &  \\[3pt]
 6 & max of $A$ (the limiting value) & $\approx 0.141064$ & 0.443166 & $\approx1.09229$ & 0.1780216 \\
 \hline \hline
 \multicolumn{5}{l}{$^*$Results from Tanaka's program} \\
\end{tabular}
\end{table*}
\begin{table*}[!]
  \caption[]{\label{tab:c} The wave speed $c$ and the mean water level $\eta_0$ for steep Stokes waves depending on their steepness $A$ ($\varepsilon=\pi A$)$^\sharp$.
  The maximal relative errors $Er_{\mathrm{max}}$ of corresponding approximate solutions demonstrate great advantage
  of fractional approximations over ordinary Fourier approximations}
\smallskip
\renewcommand{\arraystretch}{1.2}
\iftwocol{\small}{}
\begin{tabular}{llllllllc}
\hline \hline
 & &\multicolumn{3}{c}{Ordinary Fourier approximations$^1$} & \multicolumn{3}{c}{Fractional Fourier approximations$^2$} & \multicolumn{1}{c}{Tanaka's program}\\
 \multicolumn{1}{c}{$A$} & \multicolumn{1}{c}{$\varepsilon$} & \multicolumn{1}{c}{$c$} &
 \multicolumn{1}{c}{$\eta_0$} & \multicolumn{1}{c}{$Er_{\mathrm{max}},~\%$} & \multicolumn{1}{c}{$c$} & \multicolumn{1}{c}{$\eta_0$} & \multicolumn{1}{c}{$Er_{\mathrm{max}},~\%$} & \multicolumn{1}{c}{$c$} \\
 \hline
 0.14 & $\approx 0.439823$ & 1.0926149034 & $-2.8\cdot10^{-20}$ & $3.8\cdot10^{-10}$ & 1.0926149034$^a$ & $-2.4\cdot10^{-40}$ & $2.9\cdot10^{-20}$ & 1.0926149034 \\[3pt]
 0.1406 & $\approx 0.441708$ & 1.0923377398 & $-5.1\cdot10^{-13}$ & $1.6\cdot10^{-5}$ & 1.0923377499$^a$ & $-1.1\cdot10^{-22}$ & $3.2\cdot10^{-11}$ & 1.0923377499 \\[3pt]
 \textbf{0.14092} & $\approx 0.442713$ & 1.09227614 & $-1.9\cdot10^{-9}$ & $2.9\cdot10^{-3}$ & \textbf{1.0922768392} & $-2.0\cdot10^{-14}$ & $2.1\cdot10^{-6}$ & \textbf{1.0922768392} \\[3pt]
 0.141 & $\approx 0.442965$ & 1.0922815 & $-1.3\cdot10^{-8}$ & $9.6\cdot10^{-3}$ & 1.0922809 & $-1.7\cdot10^{-11}$ & $1.2\cdot10^{-4}$ & 1.0922808596 \\[3pt]
 0.14103 & $\approx 0.443059$ & 1.0922875 & $-2.8\cdot10^{-8}$ & $1.9\cdot10^{-2}$ & 1.0922841 & $-2.2\cdot10^{-10}$ & $5.8\cdot10^{-4}$ & 1.0922836847 \\[3pt]
 \textbf{0.141056} & $\approx 0.443141$ & 1.0922966 & $-6.1\cdot10^{-8}$ & $2.4\cdot10^{-2}$ & 1.0922877 & $-2.6\cdot10^{-9}$ & $2.5\cdot10^{-3}$ & \textbf{1.0922851495} \\[3pt]
 0.14106 & $\approx 0.443153$ & 1.0922987 & $-7.0\cdot10^{-8}$ & $2.6\cdot10^{-2}$ & 1.0922886 & $-4.2\cdot10^{-9}$ & $3.3\cdot10^{-3}$ & 1.0922851047 \\
 & & & & & 1.0922871$^b$ & $-2.3\cdot10^{-9}$ & $2.2\cdot10^{-3}$ & \\[3pt]
 0.141064 & $\approx 0.443166$ & 1.0923011 & $-8.1\cdot10^{-8}$ & $2.8\cdot10^{-2}$ & 1.0922902 & $-9.0\cdot10^{-9}$ & $5.2\cdot10^{-3}$ & \multicolumn{1}{c}{---$^*$} \\
\hline \hline
 & & \multicolumn{3}{l}{$^1 N=250,~M=4N$} & \multicolumn{4}{l}{$^2 N=120,~M=2.5N,~y_0=0.9,~\gamma=0.92$} \\
 & & \multicolumn{3}{l}{$^\sharp$The extremums in wave speed are bold-faced} & \multicolumn{4}{l}{$^a N=120,~M=2N,~~~~y_0=1,~~~~\gamma=0.9$} \\
 & & & & & \multicolumn{4}{l}{$^b N=150,~M=2.5N,~y_0=0.9,~\gamma=0.92$} \\
 & & & & & \multicolumn{4}{l}{$^*$The maximum steepness in Tanaka's program is $A\approx0.1410635$} \\
\end{tabular}
\end{table*}

\subsection{\textit{Physical quantities}} \label{sec:phys}
Once the coefficients $\alpha_n$, $\eta^{(\gamma)}_n$ and the wave
phase speed $c$ are found, a variety of wave characteristics can
be calculated. The velocity potential, stream function, and the
horizontal and vertical velocities of fluid particles are as
follows:
\begin{eqnarray}
\Phi(\theta,~y)&=&2c\sum_{n=1}^N \alpha_n\; \mathrm{Im}
\bigl(T^n(\theta,~y)\bigr); \label{eq:Phi}\\
\Psi(\theta,~y)&=&2c\sum_{n=1}^N \alpha_n\; \mathrm{Re}
\bigl(T^n(\theta,~y)\bigr); \label{eq:Psi}\\
\Phi_{\theta}(\theta,~y)&=&2c\sum_{n=1}^{N+1} \beta_n\;
\mathrm{Re}\bigl(T^n(\theta,~y)\bigr); \label{eq:Phi_x}\\
\Phi_y(\theta,~y)&=&2c\sum_{n=1}^{N+1} \beta_n\; \mathrm{Im}
\bigl(T^n(\theta,~y)\bigr); \label{eq:Phi_y}\\
\beta_n &=& n\alpha_n - (n-1)\alpha_{n-1} \exp{(-y_0)}. \nonumber
\end{eqnarray}

The horizontal and vertical accelerations of fluid particles are
as follows:
\begin{eqnarray}
 \frac{\mathrm{d}^2 \theta}{\mathrm{d} t^2}&=&\Phi_{\theta\theta}\left(\Phi_\theta-c\right)+\Phi_{y\theta}\Phi_y, \nonumber \\
 \frac{\mathrm{d}^2y}{\mathrm{d}t^2}&=&\Phi_{y\theta}\left(\Phi_\theta-c\right)-\Phi_{\theta\theta}\Phi_y; \label{eq:accel}
\end{eqnarray}
where
\begin{eqnarray}
\Phi_{\theta\theta}(\theta,~y)&=&-2c\sum_{n=1}^{N+2} \mu_n\;
\mathrm{Im}\bigl(T^n(\theta,~y)\bigr); \label{eq:Phi_xx}\\
\Phi_{y\theta}(\theta,~y)&=&\phantom{-}2c\sum_{n=1}^{N+2} \mu_n\;
\mathrm{Re} \bigl(T^n(\theta,~y)\bigr); \label{eq:Phi_yx}\\
\mu_n &=& n\beta_n - (n-1)\beta_{n-1} \exp{(-y_0)}. \nonumber
\end{eqnarray}

The Stokes flow $K$, the wave impulse $I$, and the wave kinetic
energy $E_{\mathrm{Kin}}$ are as follows (see \citet{Cokelet}):
\begin{equation}\label{eq:K+I+Kin}
  K=B+\eta_0,~~I=K c, ~~E_{\mathrm{Kin}}=c I/2;
\end{equation}
where $\eta_0=\overline{\eta(\theta)}$ is the wave mean level and
is determined from relation (\ref{eq:Bessel}).

The wave potential energy $U$ is calculated as follows:
\begin{eqnarray} \label{eq:U}
  \lefteqn{U = \frac{1}{2\pi}\int\limits_{0}^{2\pi}\frac{1}{2}\;
  \eta^2(\chi)\;\mathrm{d}\theta=
  \frac{1}{2}\bigl(\eta^{(\gamma)}_0\bigr)^2+\sum_{n_1=1}^{M}\bigl(\eta^{(\gamma)}_{n_1}\bigr)^2-} \\
  \lefteqn{\frac{\gamma}{2}\Bigl(\eta^{(\gamma)}_0\,\eta^{(\gamma)}_1+
  \sum_{n_1=1}^{M}\eta^{(\gamma)}_{n_1}\bigl(\eta^{(\gamma)}_{n_1-1}+\eta^{(\gamma)}_{n_1+1}\bigr)\Bigr).} \nonumber
\end{eqnarray}

\section{Regular and irregular flows} \label{sec:results}
\subsection{\textit{Stokes waves}} \label{sec:Stokes}
\begin{figure*}[!]
\begin{center}
\includegraphics[width=10.5cm]{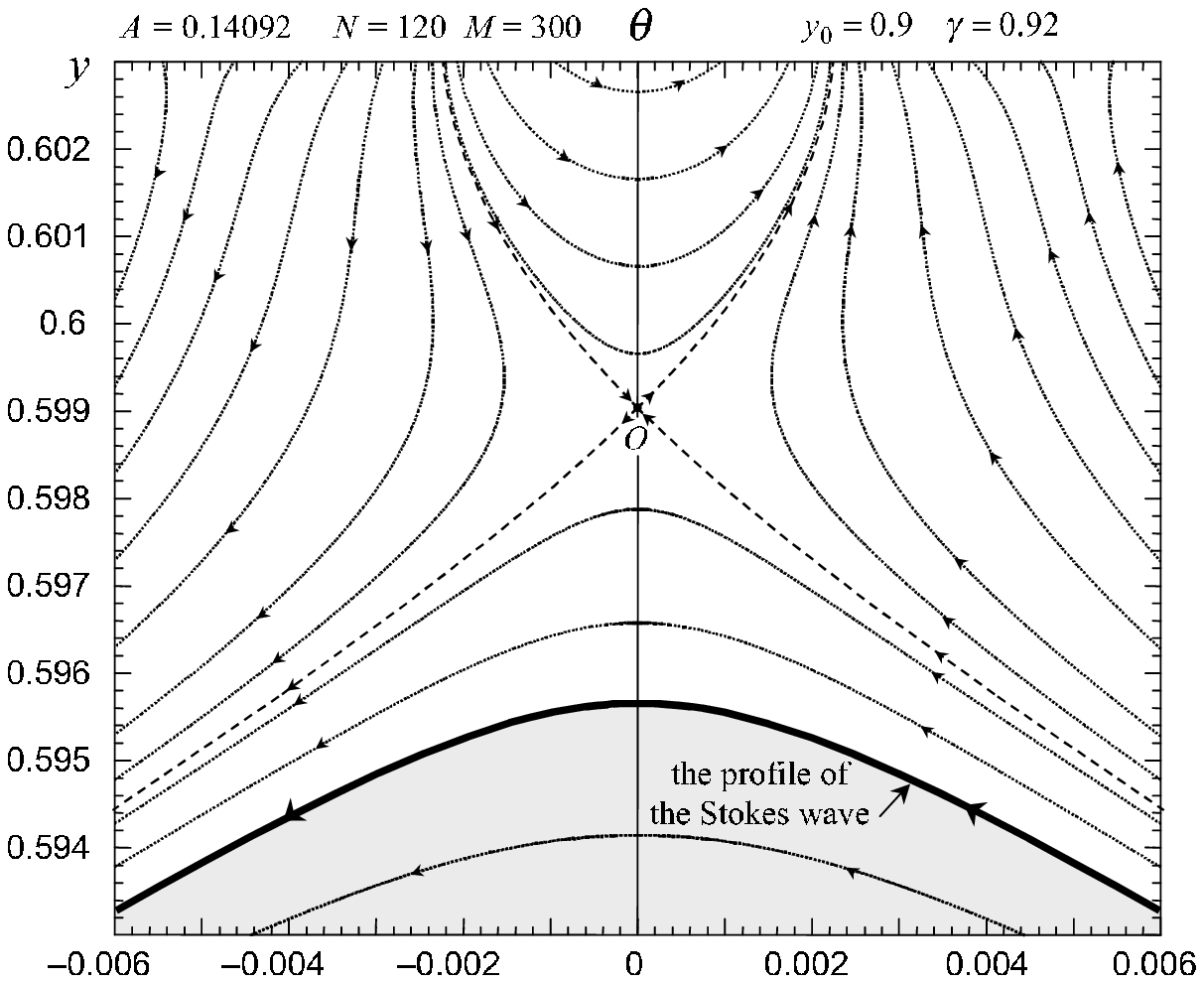}
\caption{\label{fig:regular} The streamlines of the almost highest
regular flow near the wave crest, mapped outside the domain filled
by fluid, in the wave related frame of reference.}
\end{center}
\includegraphics[width=17.5cm]{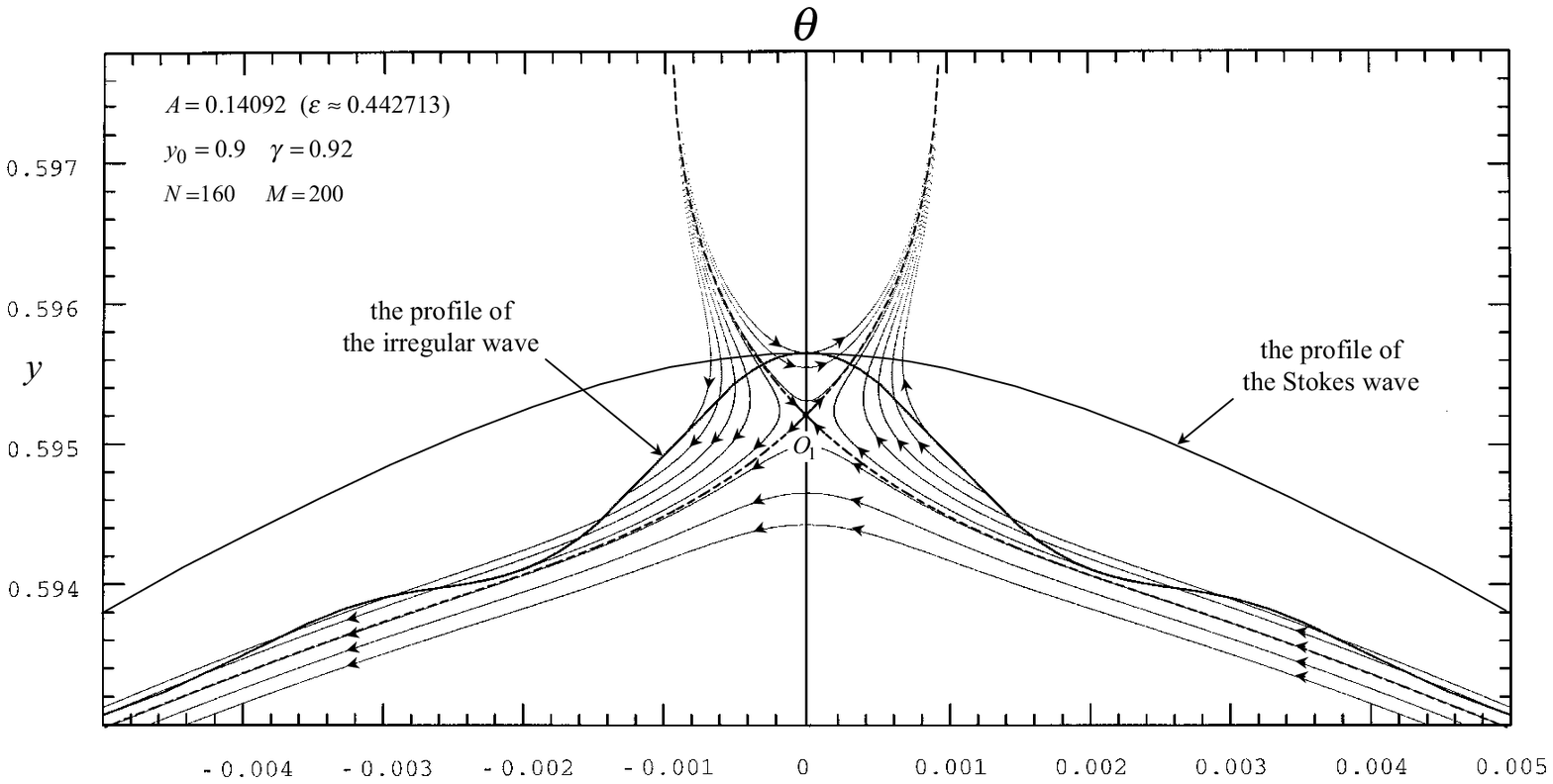}
\caption{\label{fig:irregular} The streamlines of the almost
highest irregular flow near the wave crest, in the wave related
frame of reference. The profile of the Stokes wave of the same
steepness is also included for comparison.}
\end{figure*}

The dependence $c(A)$ of the phase speed of almost highest Stokes
waves on their steepness calculated using fractional Fourier
approximations is shown in Fig.~\ref{fig:c(A)} by the branch
1-2-3-6. And the corresponding dependence $c(\eta_1)$ of the phase
speed on the first harmonic of the elevation is presented in
Fig.~\ref{fig:c(eta1)} by the branch 1-2-3-4-6. The corresponding
points of extremums in phase speed $c$, the first harmonic
$\eta_1$, and steepness $A$ are presented in
Table~\ref{tab:extremum}.

The advantage of fractional Fourier approximations over ordinary
Fourier approximations is well seen from Table~\ref{tab:c}. There,
the values of the wave phase speed $c$ and the mean water level
$\eta_0$ calculated using these two approaches are presented at
different values of the wave steepness $A$ up to the limiting one.
The precise value of the mean level $\eta_0$ should be zero for
infinite depth independently of steepness. Therefore, the
approximate values of $\eta_0$ can be used to estimate the
precision of approximate results. The maximal relative errors
$Er_{\mathrm{max}}$ of corresponding approximate solutions are
also presented for analysis. One can see from the relative errors
that the benefit from using fractional Fourier approximations with
parameters chosen varies from ten decimal orders for $A=0.14$
($\approx99.25\%$ of the limiting steepness) to one decimal order
for $A=0.141064$ (almost the limiting steepness). After taking
into account that a run using fractional Fourier approximations
with $N=120$, $M=2.5N$ needs approximately $2.5$ times lesser
computer time and approximately $10$ times lesser computer memory
than a run using ordinary Fourier approximations with
$N=250,~M=4N$, the advantage of fractional approximations becomes
doubtless.

Nevertheless, the results presented for the values of steepness
beyond the first minimum of $c$ ($A=0.14092$) are still less
accurate than the ones obtained from Tanaka's program, which are
also included into Table~\ref{tab:c} for comparison. This is also
well seen from Fig.~\ref{fig:c(A)}, where the second maximum of
$c$ ($A=0.141056$, the point 5) was obtained only using Tanaka's
program. The fractional Fourier approximation at $y_0=0.9$ is
sufficient to obtain the second maximum of $\eta_1$ (the point 4
in Fig.~\ref{fig:c(eta1)}), but not sufficient to trace the second
maximum of $c$. The reason is that the value $y_0=0.9$ used is
optimal for the steepness corresponding to the first minimum of
$c$, yet lesser $y_0$ being necessary for greater $A$ to improve
the precision of fractional Fourier approximations. However, using
the present computer realization of the method the authors failed
to accomplish this task due to unsatisfactory convergence of their
numerical algorithm for $y_0<0.9$. If this problem could be
resolved fractional Fourier approximations in the physical plane
would have a potential to become almost as effective as Tanaka's
method in the inverse plane.

\subsection{\textit{Irregular flows}} \label{sec:irregular}
\begin{figure*}[!]
\includegraphics[width=13.5cm]{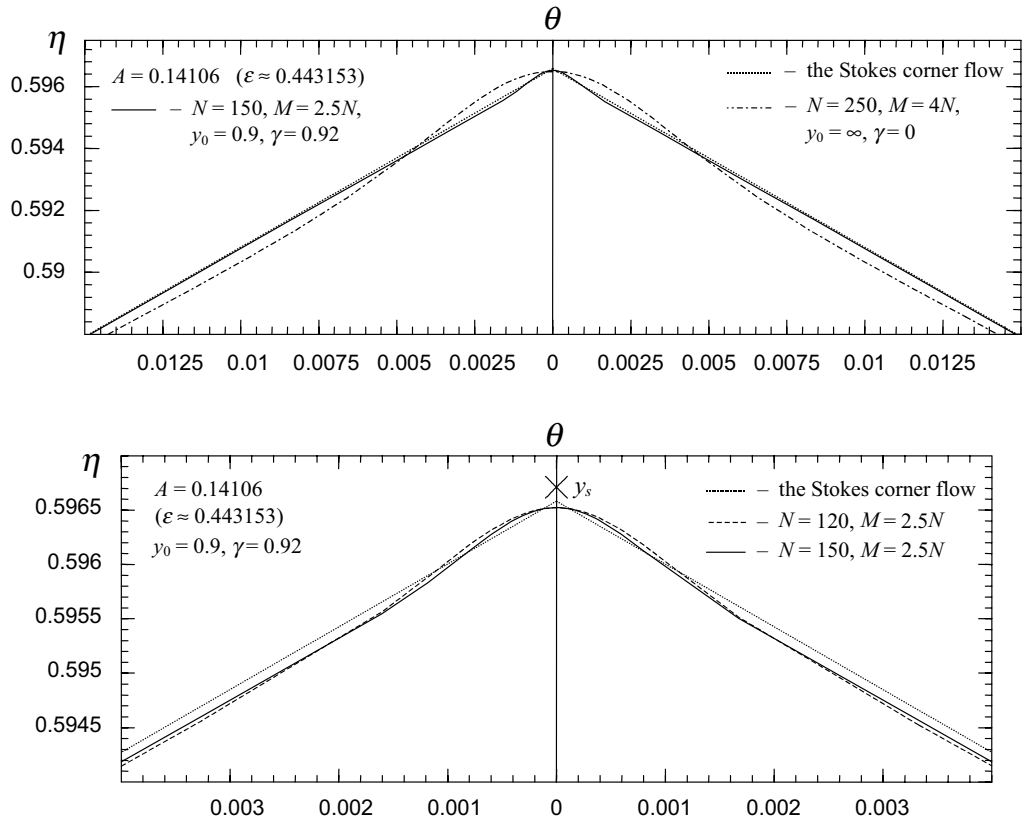}
\caption{\label{fig:Gibbs_Stokes} The Gibbs phenomenon in the
approximations to the regular Stokes wave very close to the Stokes
corner flow.}
\end{figure*}
\begin{figure*}[!]
\includegraphics[width=13.5cm]{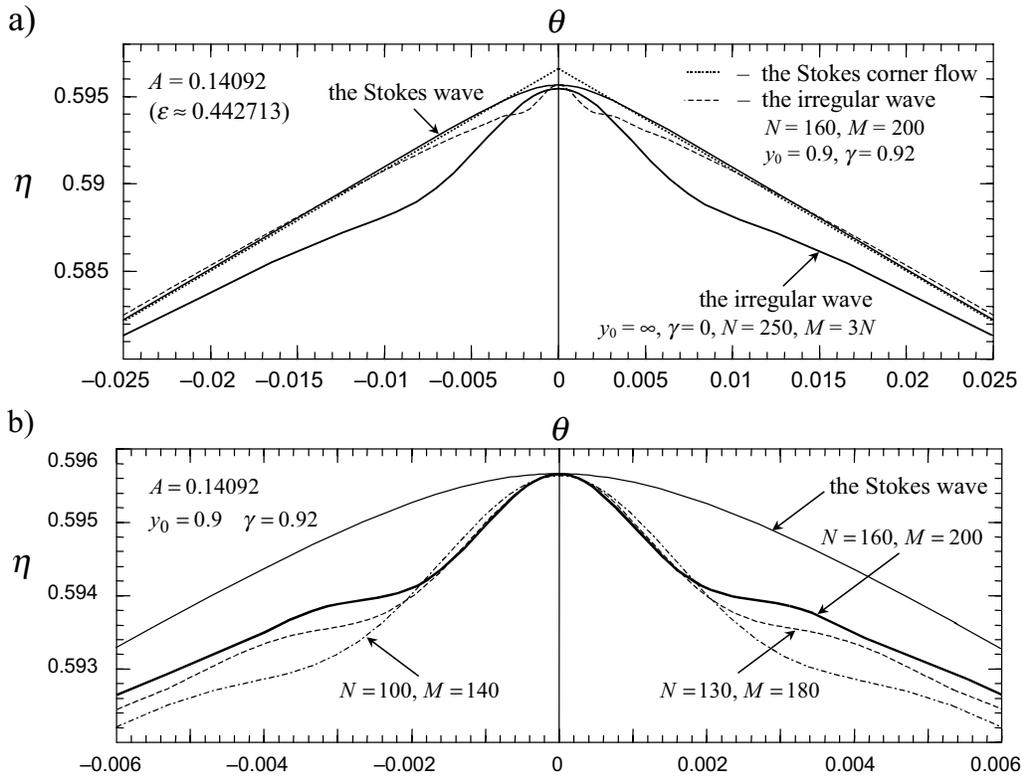}
\caption{\label{fig:spikes} a) The behavior of the Gibbs
phenomenon in the profile of the irregular wave when proceeding
from the ordinary \\ to fractional Fourier approximations. b) The
profile of the irregular wave obtained using fractional Fourier expansions \\
at different truncation numbers. The profile of the Stokes wave of
the same steepness is also included for comparison.}
\end{figure*}
\begin{table*}[!]
  \caption[]{\label{tab:irregular} The parameters of the irregular flow at $A=0.14092$ for different approximations:
  $c$ is the wave phase speed; $\eta_1$ is the first harmonic of the elevation;
  $\eta_0$ is the mean water level; $Er_{\mathrm{max}}$ is the maximal relative error of an approximate solution;
  $q(0)-c$ is the velocity at the crest in the wave related frame of reference;
  $\eta(0)$ is the height of the crest above the still water level; $y_s$ is the vertical position of the stagnation point;
  $\eta(0)-y_s$ is the distance of the stagnation point from the wave crest;
  $\alpha_{\min}$ is the minimal slope in the Gibbs oscillations of the wave profile.
  The same parameters of the regular flow at $A=0.14092$ are also included for comparison$^*$.}
\smallskip
\renewcommand{\arraystretch}{1.2}
\iftwocol{\small}{}
\begin{tabular}{cccccccccc}
\hline \hline
 \multicolumn{1}{c}{approximation} & \multicolumn{1}{c}{$c$} & \multicolumn{1}{c}{$\eta_1$} & \multicolumn{1}{c}{$\eta_0$} & \multicolumn{1}{c}{$Er_{\mathrm{max}},~\%$} & \multicolumn{1}{c}{$q(0)-c$}&
 \multicolumn{1}{c}{$\eta(0)$} & \multicolumn{1}{c}{$y_s$} & \multicolumn{1}{c}{$\eta(0)-y_s$} & \multicolumn{1}{c}{$\alpha_{\min},~^{\circ}$} \\
 \hline
 \multicolumn{10}{c}{irregular flow} \\ \hline
 $N=250,~M=3N~^1$ & 1.092427 & 0.178169 & $-1.7\cdot10^{-6}$ & $1.5\cdot10^{-1}$ & 0.0531 & 0.595445 & 0.593830 & 0.001615 & 19.7 \\[3pt]
 $N=100,~M=140~^2$ & 1.092350 & 0.178039 & $-3.9\cdot10^{-7}$ & $4.2\cdot10^{-2}$ & 0.0470 & 0.595636 & 0.594947 & 0.000689 & 12.1 \\[3pt]
 $N=130,~M=180~^2$ & 1.092330 & 0.178024 & $-2.8\cdot10^{-7}$ & $3.1\cdot10^{-2}$ & 0.0455 & 0.595652 & 0.595117 & 0.000535 & 9.6 \\[3pt]
 $N=160,~M=200~^2$ & 1.092318 & 0.178017 & $-2.2\cdot10^{-7}$ & $2.4\cdot10^{-2}$ & 0.0448 & 0.595661 & 0.595221 & 0.000440 & 7.7 \\ \hline
 \multicolumn{10}{c}{regular flow} \\ \hline
 $N=120,~M=2.5N~^2$ & 1.092277 & 0.178010 & $-2.0\cdot10^{-14}$ & $2.1\cdot10^{-6}$ & $-0.0419$ & 0.595657 & 0.599019 & $-0.003362$ & --- \\
 \hline \hline
 \multicolumn{10}{l}{$^1$ ordinary Fourier approximations} \\
 \multicolumn{10}{l}{$^2$ fractional Fourier approximations $y_0=0.9,~\gamma=0.92$} \\
 \multicolumn{10}{l}{$^*$ for the Stokes corner flow, $c \approx 1.0923$, $\eta(0)=c^2/2 \approx 0.59656$, $y_s=\eta(0)$}
\end{tabular}
\end{table*}

The flow field in Stokes waves is regular, that is, fluid
particles move slower than the wave itself all over the flow
domain. In the Stokes corner flow only, the fluid particle at the
wave crest moves with velocity equal to the wave phase speed and,
therefore, is motionless with respect to the wave. Because of this
such points in the flow field are called the stagnation points.
For all the Stokes waves other than the limiting one, a stagnation
point is located outside the flow domain, as was at first shown by
\citet{Grant}. The example of such a regular flow mapped outside
the domain filled by fluid is presented in Fig.~\ref{fig:regular}.
The streamlines at the stagnation point $O$ meet at right angles
in accordance with the results of \citet{Grant} and
\citet{Long:al-h1978}.

\citet{prl2002} have numerically revealed another type of waves
with stagnation point located inside the flow domain. This
stagnation point makes streamlines be discontinuous in the
vicinity of the wave crest and form two symmetric upward and
downward jets. Because of this such waves and flows were called
irregular. The example of such an irregular flow calculated using
fractional Fourier approximations is shown in
Fig.~\ref{fig:irregular}. Note that the streamlines at the
stagnation point $O_1$ also meet at right angles. This is the
general rule valid for all the irregular and regular flows
provided that the stagnation point and the wave crest do not merge
(see Appendix~\ref{apex:90} for details).

One can see from Fig.~\ref{fig:irregular} that the particles from
the near-surface layer of the irregular flow are accelerated to
velocities greater than the wave phase speed when approaching the
crest. As a result, they form the upward jet emanating from the
front face of the wave. The acceleration of particles at the base
of the jet ranges from $2.5g$ at $\theta=0.001$ to $6g$ at the
wave crest. Such large accelerations of the water rising up the
front of the wave into the jet are known to occur in breaking
waves \citep[see][]{Peregrine:review}, where the typical maxima in
accelerations obtained from detailed unsteady numerical
computations were reported to be around $5g$. In view of these
points, irregular flows produce a simplified model for describing
the initial stage of wave breaking or, to be more precise, the
formation of spilling breakers when a localized jet is formed near
the crest following by generating whitecaps.

Irregular wave is an approximate stationary and obviously unstable
solution to the equations of motion. Although we did not
investigate its unsteady evolution, the following results are
expected. The particles with velocities greater than the wave
speed will all be jetted out away from the fluid and the crest
will break if their is no external influx into the flow domain
from the left downward jet. After finishing this non-stationary
process the flow will become regular and of lesser steepness. This
resembles the recurrence phenomenon observed by
\citet{Long:al-h1997} when computing unsteady non-linear
development of the crest instabilities of almost highest Stokes
waves resulting in a smooth transition to a periodic wave of lower
amplitude. The appearance of irregular flows may give an
explanation to this phenomenon.

Irregular flows at fixed truncation numbers $N$ and $M$ depend on
the wave steepness $A$ in the following way. As wave steepness
decreases from the limiting value, the stagnation point settles
down deeper into the flow domain and the distance between the
upward and downward jets increases. The corresponding dependence
of the phase speed of irregular waves on their steepness $c(A)$ is
shown in Fig.~\ref{fig:c(A)} by the branch 6-8. And the dependence
$c(\eta_1)$ of the phase speed on the first harmonic of the
elevation is presented in Fig.~\ref{fig:c(eta1)} by the branch
6-7-8. While moving along the irregular branch away from the
limiting point 6 the accuracy of approximate solutions at fixed
$N$ and $M$ drops. Because of this the branches corresponding to
irregular flows in Figs.~\ref{fig:c(A)}, \ref{fig:c(eta1)} have
not yet stabilized with respect to increasing the truncation
numbers $N$ and $M$, although fractional approximations
(\ref{eq:R_y0}) in combination with non-linear transformation
(\ref{eq:gamma}) are up to one decimal order more accurate than
ordinary Fourier approximations (\ref{eq:Fourier}) when
calculating irregular waves. The loop in Fig.~\ref{fig:c(A)} still
enlarges with increasing $N$, the cross-section point with the
Stokes branch moving to the left. On the contrary, the irregular
branch 6-7-8 in the dependence $c(\eta_1)$ (see
Fig.~\ref{fig:c(eta1)}) approaches to the regular branch 1-2-3-4-6
as accuracy is increased. They seem to coincide completely at
$N\rightarrow\infty$. This may explain why irregular waves were
not traced from stability analysis of Stokes waves. Similar to
almost limiting Stokes waves, lesser $y_0$ are necessary to
improve the precision of fractional Fourier approximations and to
stabilize the position of the irregular branch.

Thus, our main concern is that irregular flow is an approximate
solution, whose accuracy is not sufficient enough to make definite
conclusions even when using fractional approximations. The only
reasonable assumption is that irregular waves correspond to
singular solutions of the equations of motion. One can see from
Fig.~\ref{fig:irregular} that although streamlines of the
irregular flow are discontinuous near the wave crest the wave
profile $\eta(\theta)$ remains to be a continuous function
everywhere. This is due to the Fourier series
(\ref{eq:Fourier_Eta_gamma}) for $\eta(\theta)$ being a
single-valued smooth function, which represents an integral
characteristic of the flow. On the contrary, the stream function
$\Psi(\theta,~y)$ represents a local characteristic of the flow
and is not obligatory a single-valued dependence. Because of this
streamlines describe the discontinuous flow near the wave crest
more adequately. It is seen from Fig.~\ref{fig:irregular} that the
profile of the irregular flow oscillates while approaching the
wave crest, where a prominent peak (an overshoot) forms. This
highly resembles the Gibbs phenomenon when (i) a discontinuous
function or (ii) a continuous function with discontinuous
derivatives are approximated by a truncated set of continuous
functions \citep[see, e.g.,][]{MathMeth}. In both cases, the Gibbs
phenomenon can be used as an excellent indicator of a singularity.

The example corresponding to the case (i) is given in Appendix
\ref{apex:Gibbs}, where truncated Fourier series of a function
with infinite discontinuity are demonstrated to exhibit typical
Gibbs oscillations and an overshoot. For the case (ii), the
limiting Stokes wave can be used as an example because of the
sharp $120^\circ$ corner at the crest corresponding to a
discontinuous first derivative. Moreover, even regular Stokes
waves very close to the Stokes corner flow also exhibit Gibbs
oscillations when being approximated numerically, as was revealed
by \citet{Nekrasov_Gibbs} from the analysis of Nekrasov's integral
equation. The similar example obtained using fractional and
ordinary Fourier expansions is presented in
Fig.~\ref{fig:Gibbs_Stokes} for the Stokes wave at $A=0.14106$
($\approx99.9975\%$ of the limiting steepness
$A\approx0.1410635$). The Gibbs phenomenon is distinctly observed
even though the exact solution is known to be not sharp-crested.
The possible reason is that it has discontinuous derivatives
higher than the first one being continuous. One can notice from
Fig.~\ref{fig:Gibbs_Stokes} that the overshoot shrinks both in
vertical and horizontal scales with increasing the accuracy of
approximations in contrast to the example with infinite
discontinuity, where the height of the overshoot remains almost
constant (see Fig.~\ref{fig:Gibbs}).

Now let us analyse how an approximate irregular flow at fixed
steepness depends on improving its numerical accuracy and make a
comparison with two examples considered above. In
Fig.~\ref{fig:spikes}, the profiles of the irregular flow at
$A=0.14092$ are plotted for several approximations obtained using
fractional and ordinary Fourier expansions, the corresponding flow
parameters being presented in Table~\ref{tab:irregular}. The
following aspects can be emphasized.

1) The overshoot shrinks both in vertical and horizontal scales as
the accuracy of expansions is improved, in the same way as for the
almost limiting Stokes wave in Fig.~\ref{fig:Gibbs_Stokes}. The
horizontal and vertical dimensions of the overshoot both become
approximately five times as small when using more accurate
fractional approximations instead of ordinary Fourier
approximations. And the distance $\eta(0)-y_s$ between the wave
crest and the stagnation point becomes approximately four times as
small. This correlates with decreasing the maximal relative error
$Er_{\max}$ of the solutions approximately by a factor of six. The
conclusion can be made that the overshoot is likely to shrink into
a single point when increasing accuracy further similar to the
case of the almost limiting Stokes wave.

2) The overshoot has an almost fixed vertical position in contrast
to the example with infinite discontinuity in
Fig.~\ref{fig:Gibbs}, where the overshoot moves to infinity as
accuracy is improved. This indicates that if the profiles of
irregular flows are indeed discontinuous they seem to exhibit a
finite discontinuity. The wave height $\eta(0)$ quickly stabilizes
with increasing accuracy and seems not to tend to the height of
the Stokes corner flow.

3) The particle velocity at the crest $q(0)-c$ in the wave related
frame of reference decreases but much less rapidly than the
relative error $Er_{\max}$ of the corresponding solutions.

4) The minimal slope $\alpha_{\min}$ in the Gibbs oscillations of
the wave profile decreases, the oscillatory tails approaching the
horizontal line. This may be the indication that the profile has
two symmetric finite discontinuities in the vicinity of the wave
crest.

These observations and the data presented do not give the final
understanding of irregular flows at $N\rightarrow\infty$. At
present, we can only guess what exact solutions to the equations
of motion irregular waves do approximate. We admit the following
two alternatives.

(i) The particle velocity at the crest $q(0)-c$ drops to zero, the
overshoot shrinks into a single point, and the stagnation point
merges with the wave crest. {\it Then} irregular flows approximate
a family of sharp-crested corner flows similar to the Stokes
corner flow but of lesser steepness.

(ii) The particle velocity at the crest $q(0)-c$ does not tend to
zero and the stagnation point remains in the flow domain. {\it
Then} irregular flows are indeed discontinuous and their profiles
are not single-valued and have two symmetric finite
discontinuities in the vicinity of the wave crest.


Further investigation is necessary to single out one of these
possibilities or probably to come to some other conclusion.

\section{Conclusions} \label{sec:concl}
Fractional Fourier approximations for the velocity potential in
combination with non-linear transformation of the horizontal
scale, which concentrates a numerical emphasis on the crest
region, turned out to be much more efficient than ordinary Fourier
approximations when computing both steep regular and irregular
flows. Nevertheless, further improvement of the numerical
algorithm is necessary to achieve the accuracy of Tanaka's method
when calculating almost limiting Stokes waves and to attain the
final understanding of irregular flows. One of the possible ways
is to use the following multi-term fractional expansion with
several different parameters $y_k$:
$$
R(\theta,~y;~\{y_k\}) = \sum_{k=0}^K \sum_{n=1}^{N_k}
\frac{\alpha^{(k)}_n}{\bigl(\exp(-y_k) - \exp \left(- y - i\theta
\right)\bigr)^n}.
$$

Although the proposed approach was formulated in the framework of
the canonical model for infinite depth, its practical application
is much broader. \citet{abstr:EGS03_ST} successfully employed
fractional approximations for computing gravity-capillary waves.
When $y_0$ is located inside the flow domain fractional
approximations may be applied for calculating vortex structures
and solitary waves. The latter possibility was realized by
\citet{Clamond:rKdV} using his renormalized cnoidal wave
approximation (the first term of the fractional Fourier
approximation). He computed an algebraic solitary wave on deep
water and traced how it changes after taking into account surface
tension (it is not known, however, how this algebraic solution
depends on picking up higher terms of the fractional expansion).
In the case of finite depth $h$, the fractional Fourier expansion
will be as follows:
\begin{eqnarray}
\lefteqn{R(\theta,~y;~y_0) = \sum_{n=1}^{N}\Bigl(
\frac{\alpha^{(+)}_n}{\bigl(\exp(-y_0) - \exp \left(- y - h -
i\theta \right)\bigr)^n}- \phantom{\Bigr)}} \nonumber \\
\lefteqn{\phantom{R(\theta,~y;~y_0) = \sum_{n=1}^{N}\Bigl(}
\frac{\alpha^{(-)}_n}{\bigl(\exp(-y_0) - \exp \left(y + h +
i\theta \right)\bigr)^n}\Bigr).} \nonumber
\end{eqnarray}
Finally, fractional expansions may also be generalized to the case
of 3D waves and non-ideal fluid.

Fractional approximations allowed us to gain more detailed
knowledge about the properties of irregular flows. Irregular waves
were proved to correspond to singular solutions of the equations
of motion. Because of this their existence does not contradict to
the uniqueness theorem of \citet{Garabedian} since it deals with
regular continuous solutions only. The following two alternatives
for the exact solutions associated with approximate irregular
flows were singled out: (i) the profile of an exact solution is
continuous with sharp corner at the crest (a discontinuous first
derivative) similar to the limiting Stokes wave but is of lesser
steepness; (ii) the profile of an exact solution has two symmetric
finite discontinuities in the vicinity of the wave crest and is
not single-valued. Further analysis, however, should be carried
out to make a final conclusion. One of the possible ways is to
investigate how an approximate irregular flow depends on taking
into account surface tension and to make a comparison with new
limiting forms for gravity-capillary waves recently obtained by
\citet{Kharif_ST:1996}.

To conclude note that the formation of jets from irregular
progressive waves resembles the occurence of vertical jets with
sharp-pointed tips from standing gravity waves forced beyond the
maximum height, as has recently been reported by
\citet{Long:jets2001}.

\balance 

\appendix

\section{The relations between the Fourier coefficients in the $\theta$- and $\chi$-spaces} \label{apex:gamma}
Taking into account nonlinear transformation (\ref{eq:gamma}) the
coefficients in Fourier series (\ref{eq:Fourier_Eta}) and
(\ref{eq:Fourier_Eta_gamma}) are connected as follows
\begin{eqnarray}
\eta_0&=&\eta^{(\gamma)}_0-\gamma \eta^{(\gamma)}_1, \nonumber \\
\eta_n&=&\frac{1}{n}\sum_{n_1=1}^{M}n_1\eta^{(\gamma)}_{n_1}\bigl(J_{n-n_1}(n\gamma)-J_{n+n_1}(n\gamma)\bigr),
\label{eq:Bessel}
\end{eqnarray}
$n=\overline{1,~\infty}$; $J_n(z)$ being the Bessel function of
the first kind.

\section{The Jacoby matrix} \label{apex:Jacoby}
The Jacoby matrix is composed of the coefficients at the
infinitesimal variations $\delta c$, $\delta\eta^{(\gamma)}_0$,
$\delta\eta^{(\gamma)}_{n_1}$ ($n_1=\overline{2,~M}$),
$\delta\alpha_{n_1}$ ($n_1=\overline{1,~N}$) of the unknown
variables in the following variations of equations
(\ref{eq:Dyn_n}) and (\ref{eq:Kin_n}):
\begin{eqnarray*}
\delta\mathcal{D}_n&=&\sum_{n_1=1}^N
\alpha_{n,\,n_1}^{\,22}\delta\alpha_{n_1}+\sum_{n_1=0}^M\alpha_{n,\,n_1}^{\,21}\delta\eta^{(\gamma)}_{n_1}+2c\,d_n\delta
c; \\
\delta\mathcal{K}_n&=&\sum_{n_1=1}^N
\alpha_{n,\,n_1}^{\,12}\delta\alpha_{n_1}+\sum_{n_1=0}^M\alpha_{n,\,n_1}^{\,11}\delta\eta^{(\gamma)}_{n_1};
\end{eqnarray*}
where
\begin{eqnarray*}
\alpha_{n,\,n_1}^{\,11}&=&\alpha_{n-n_1}^{\,11}+\alpha_{n+n_1}^{\,11},~\alpha_{n,\,0}^{\,11}=\alpha_{n}^{11};\\
\alpha_{n,\,n_1}^{\,12}&=&2\,t_n^{(n_1)};~~\alpha_{n}^{11}=2\sum_{n_1=1}^{N+1}\beta_{n_1}
t_n^{(n_1)}-\delta_{n,\,0};\\
\alpha_{n,\,n_1}^{\,21}&=&\alpha_{n-n_1}^{\,21}+\alpha_{n+n_1}^{\,21},~\alpha_{n,\,0}^{\,21}=\alpha_{n}^{21};\\
\alpha_{n}^{21}&=&2c^2\sum_{n_1=1}^{N+1}\beta_{n_1}\Bigl(n_1
\bigl(t_n^{(n_1)}-t_n^{(n_1+1)}e^{-y_0}\bigr)-\phantom{\Bigr)}\\
&&\kern-0.5cm\sum_{n_2=n_1}^{N+1}(2-\delta_{n_1,\,n_2})\,\beta_{n_2}\bigl((n_1+n_2)\,t_n^{(n_1,\,n_2)}-\phantom{\bigr)}\\
\phantom{\Bigl(\bigl(}&&\kern-0.45cm
n_1\,t_n^{(n_1+1,\,n_2)}e^{-y_0}-n_2\,t_n^{(n_1,\,n_2+1)}e^{-y_0}\bigr)\Bigr)-\delta_{n,\,0};\\
\alpha_{n,\,n_1}^{\,22}&=&2n_1c^2\Bigl(t_n^{(n_1)}-t_n^{(n_1+1)}e^{-y_0}-\phantom{\Bigr)}\\
\phantom{\Bigl)}&&2\sum_{n_2=1}^{N+1}\beta_{n_2}\bigl(t_n^{(n_1,\,n_2)}-t_n^{(n_1+1,\,n_2)}e^{-y_0}\bigr)\Bigr).
\end{eqnarray*}
Note that the variation $\delta\eta^{(\gamma)}_1$ should be
expressed in terms of the rest variations
$\delta\eta^{(\gamma)}_n$, $n=\overline{2,~M}$ using expression
(\ref{eq:A}) if the governing parameter is the steepness $A$ or
the relation (\ref{eq:Bessel}) at $n=1$ if the governing parameter
is the first harmonic $\eta_1$ of the elevation in the
$\theta$-space.

\section{The stagnation point} \label{apex:90}
The point in the flow field, where fluid particles are motionless
in the wave related frame of reference, is called the stagnation
point. For symmetric regular/irregular flows, the stagnation point
is located above/below the wave crest outside/inside the flow
domain on the axis $\theta=0$. Then its vertical position $y_s$ is
determined as follows
\begin{equation}\label{eq:ys}
  \Phi_{\theta}(0,~y_s)=c.
\end{equation}

To find the velocity field $\Phi_{\theta}(\theta,~y)$,
$\Phi_y(\theta,~y)$ in the infinitesimal vicinity
$\theta=\tilde{\theta}$ ($\tilde{\theta}\rightarrow 0$),
$y=y_s+\tilde{y}$ ($\tilde{y}\rightarrow 0$) of the stagnation
point it is sufficient to linearize there the expansions
(\ref{eq:Phi_x}), (\ref{eq:Phi_y}) that represent exact velocity
field at $N\rightarrow\infty$. For this, one should linearize the
functions $T^n(\theta,~y)$ around the stagnation point as follows
\begin{eqnarray}\label{eq:T_ys}
\lefteqn{~~T(\theta,~y)=T(y_s+\tilde{y}+i\tilde{\theta})=T(y_s)+T'(y_s)(\tilde{y}+i\tilde{\theta});}\nonumber\\
\lefteqn{T^n(\theta,~y)=T^n(y_s)+n
T'(y_s)T^{n-1}(y_s)(\tilde{y}+i\tilde{\theta}).}
\end{eqnarray}
Then, after taking into account condition (\ref{eq:ys}), the
Eqs.~(\ref{eq:traject}) for particle trajectories attain the
following form in the vicinity of the stagnation point:
$$
 \frac{d \tilde{\theta}}{dt}=a\tilde{y}, ~~
 \frac{d \tilde{y}}{dt}=a\tilde{\theta};~~a=2c\sum\limits_{n=1}^{\infty} n\beta_n
 T'(y_s)T^{n-1}(y_s).
$$
Therefore, the equations for the streamlines are
$\tilde{\theta}=\pm\tilde{y}$. Actually, this is a direct
consequence of the fact that any solution to the Laplace equation
(\ref{eq:Lapl}) should depend not on the variables ($\theta$, $y$)
separately but on their combination $y+i\theta$.

Thus, the streamlines meet at right angles ($90^\circ$) at the
stagnation point. This fact is valid for any flow provided that
the stagnation point and the wave crest do not merge. Otherwise,
$y_s=\eta(0)$; $\tilde{\theta}$ and $\tilde{x}$ are not
independent variables, and linearization (\ref{eq:T_ys}) is not
valid. The known example is the limiting Stokes corner, where the
streamlines turn out to meet at $120^\circ$ angle, as was at first
shown by \citet{Stokes}. The question appears then what is a
physical background for such a sharp transition from $90^\circ$ to
$120^\circ$ angle when proceeding from regular Stokes flows to the
Stokes corner flow? \citet{Grant} conjectured that a continuous
approach to the limiting Stokes wave is possible only if the
Stokes corner flow has several coalescing singularities. The
existence of irregular flows may give answer where these multiple
singularities arise from. The Stokes corner flow seems to be
formed due to merging the stagnation points of regular and
irregular flows.

\section{The Gibbs phenomenon} \label{apex:Gibbs}
\begin{figure}[h]
\includegraphics[width=8.5cm]{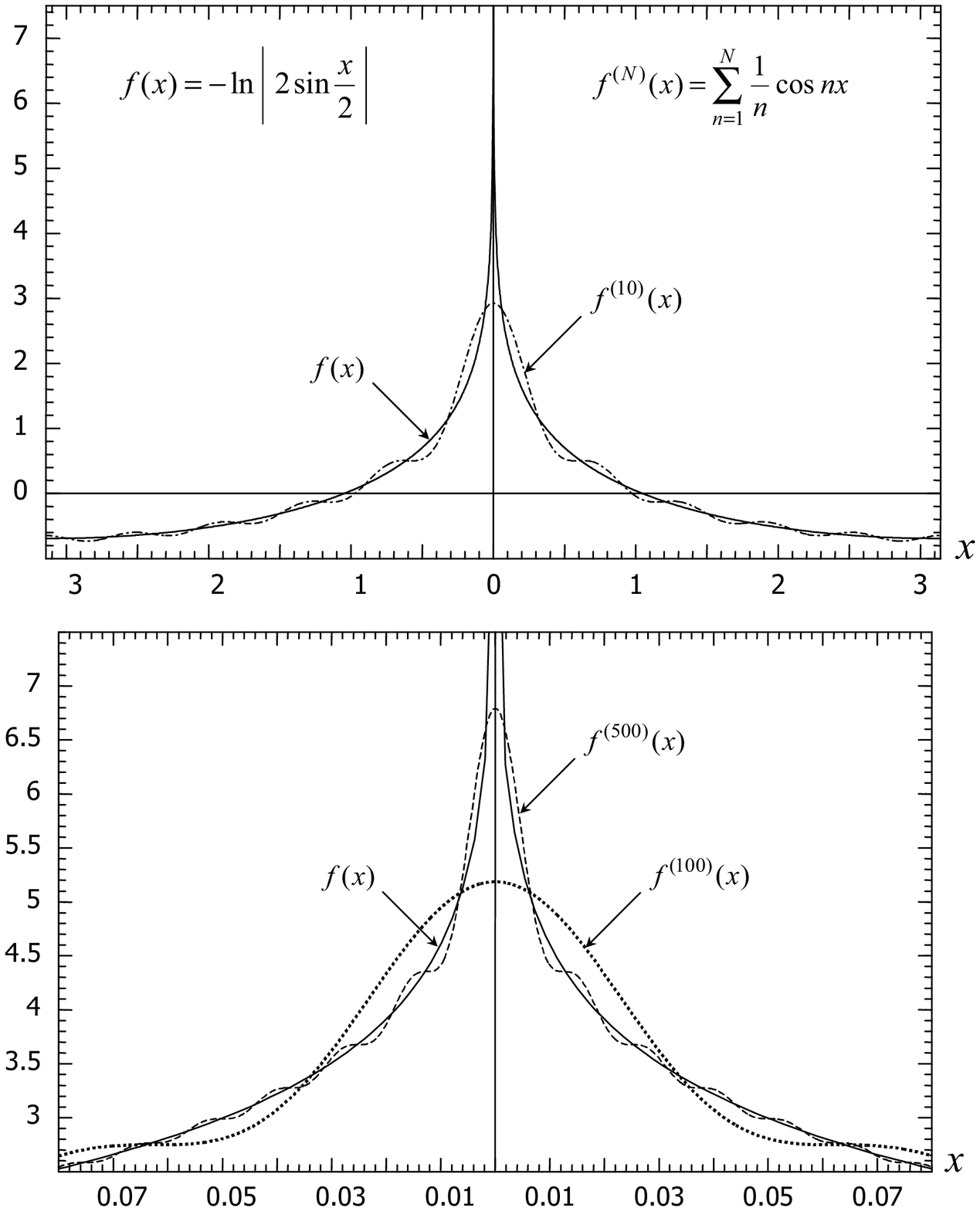}
\caption{\label{fig:Gibbs} The truncated Fourier series of the
discontinuous function. The Gibbs phenomenon.}
\end{figure}

Consider the following $2\pi$-periodic function
\begin{equation}
f(x)=-\ln\Bigl|2\sin\frac{x}{2}\Bigr|
\end{equation}
with infinite discontinuity at $x=2\pi k,~k\in \mathbb{Z}$. This
function constitutes a part of the kernel of Nekrasov's integral
equation \citep[see][]{Nekrasov_Gibbs}. The truncated Fourier
series of the function $f(x)$ have the following form
\citep[see][]{MathMeth}:
\begin{equation}
f^{(N)}(x)=\sum\limits_{n=1}^{N}\frac{1}{n}\cos(nx),~~f(x)=\lim\limits_{N\rightarrow\infty}f^{(N)}(x).
\end{equation}
In this case, the discontinuous function $f(x)$ is approximated by
the continuous functions $f^{(N)}(x)$. One can see from
Fig.~\ref{fig:Gibbs} that instead of infinite discontinuities, the
functions $f^{(N)}(x)$ have rounded peaks (overshoots) with
symmetric oscillatory tails that descend as the distance from the
point of discontinuity increases. This is the well known Gibbs
phenomenon, which always takes place when approximating
discontinuous functions by the truncated Fourier series
\citep[see][]{MathMeth}. As the number $N$ is increased, the
functions $f^{(N)}(x)$ approximate the function $f(x)$ more
precisely. The peak moves upwards and the oscillatory tails move
closer to the point of discontinuity, their amplitude and period
decreasing. Nevertheless, the height of the peak (the vertical
distance between the point $x=0$ and the point, where the
oscillatory tail initiates) remains almost constant with
increasing $N$. Because of this the truncated Fourier series
representation remains unreliable in the vicinity of a
discontinuity even for high enough $N$.

\begin{acknowledgements}
We are grateful to Prof. M. Tanaka for being so kind to place at
our disposal his program for calculating Stokes waves. We express
thanks and appreciation to Professors D.H. Peregrine, C. Kharif,
V.I. Shrira, and Dr. D. Clamond for many valuable advices and
fruitful discussions. The research of I. Gandzha has been
supported by INTAS Fellowship YSF 2001/2-114.
\end{acknowledgements}

\bibliographystyle{egs}

\end{document}